Article

# An Atomistic Insight into Moiré Reconstruction in Twisted Bilayer Graphene beyond the Magic Angle

Aditya Dey,*,§ Shoieb Ahmed Chowdhury,§ Tara Peña, Sobhit Singh, Stephen M. Wu, and Hesam Askari

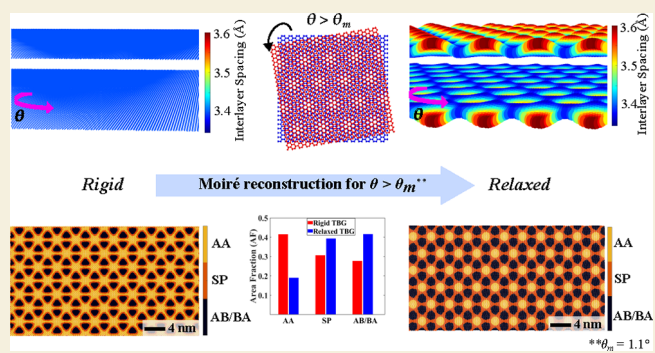

**ABSTRACT:** Twisted bilayer graphene exhibits electronic properties strongly correlated with the size and arrangement of moiré patterns. While rigid rotation of the two graphene layers results in a moiré interference pattern, local rearrangements of atoms due to interlayer van der Waals interactions result in atomic reconstruction within the moiré cells. Manipulating these patterns by controlling the twist angle and externally applied strain provides a promising route to tuning their properties. Atomic reconstruction has been extensively studied for angles close to or smaller than the magic angle ($\theta_m = 1.1°$). However, this effect has not been explored for applied strain and is believed to be negligible for high twist angles. Using interpretive and fundamental physical measurements, we use theoretical and numerical analyses to resolve atomic reconstruction in angles above $\theta_m$. In addition, we propose a method to identify local regions within moiré cells and track their evolution with strain for a range of representative high twist angles. Our results show that atomic reconstruction is actively present beyond the magic angle, and its contribution to the moiré cell evolution is significant. Our theoretical method to correlate local and global phonon behavior further validates the role of reconstruction at higher angles. Our findings provide a better understanding of moiré reconstruction in large twist angles and the evolution of moiré cells under the application of strain, which might be potentially crucial for twistronics-based applications.

## I. INTRODUCTION

Engineering two-dimensional (2D) materials by controlling the stacking orientation of atomic layers has emerged as a powerful technique to manipulate their mechanical and optoelectronic properties. Bilayer graphene (BLG) is one of the simplest van der Waals (vdW) structures that display diverse physical properties such as contrasting electronic properties that depend on the stacking arrangement.[1−4] Introducing a relative rotation between the layers forms the twisted bilayer graphene (TBG) in which the atoms create a periodic hexagonal superlattice called a "moiré pattern" (MP).[5,6] The emergence of this pattern is due to the atoms occupying different relative interlayer positions compared to BLG with a moiré cell size ($L_m$) that is inversely correlated with the twist angle ($\theta$) as $L_m = a/(2\sin(\theta/2))$ where $a$ is the lattice constant of graphene. Applying other mechanical stimuli, such as inequivalent strain to the individual layers of TBG, can further manipulate the shape of the pattern. Thus, designing both twist angle and heterostrain provides a promising route to obtain a full range of MP periodicities and symmetries for exciting optoelectronic applications.[7−9]

The atomic arrangements within MPs are influenced by the interlayer vdW forces between the 2D layers that considerably influence the atomic arrangement landscape. To manifest this influence, we can consider a hypothetical intermediate configuration where atoms are rigidly twisted in their plane. Consequently, the well-defined BLG stacking configurations of AA, AB, and SP types emerge and spatially vary throughout the superlattice.[10,11] Upon allowing atomic relaxation, an atomic-scale reconstruction occurs, and local stacked regions evolve to their true minimum local energy configuration. This process is known as atomic or moiré reconstruction.[12,13] Previous studies have reported this phenomenon for low angle TBGs, especially in the vicinity of or below the "magic angle" ($\theta_m = 1.1°$).[14,15] As the size of the MP shrinks with increasing $\theta$ and leaves less space for reconfiguration of atoms, experimental observation of moiré reconstruction becomes a challenge and is generally assumed to be absent for $\theta > 2°$.[14,16,17] Since large angle TBGs contain the same atomic registry as small twist angles, it is

A

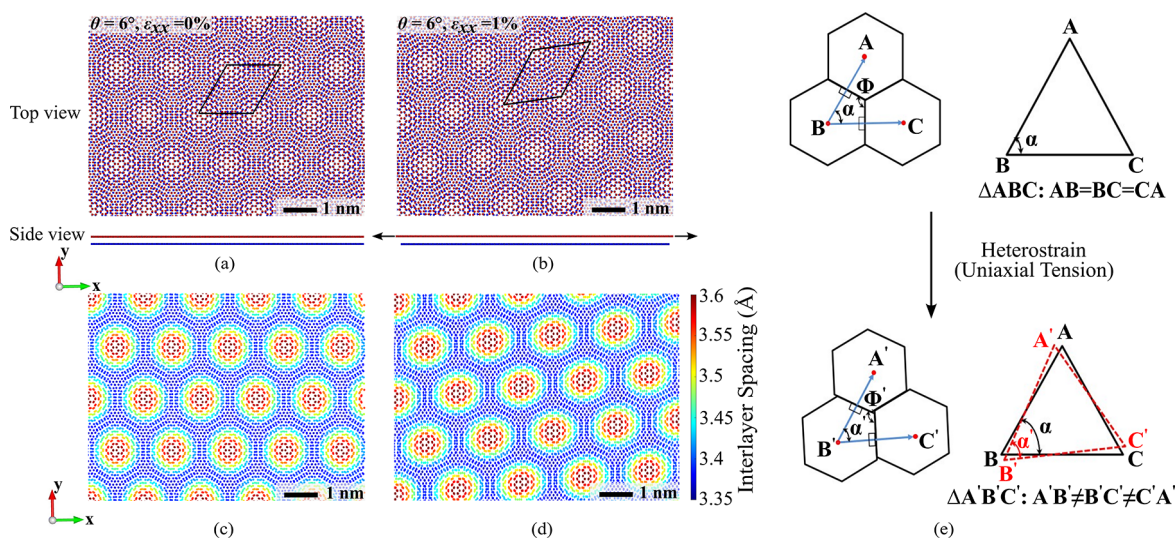

Figure 1. Atomistic model. Relaxed atomistic structures illustrate how the periodic moiré superlattice is formed and how its shape evolves with strain (a and b). Arrows in the side view show the direction of strain. Unlike BLG, where a single interlayer distancing is expected, a twist results in spatial variations of interlayer distancing as shown for (c) unstrained and (d) strained TBGs. Data presented for the twist angle of $\theta = 6°$ and uniaxial strain of 1%. Scale bars for the real space lattice and contour plots are shown with thick black lines. (e) Real space geometric analysis demonstrating the distortion of MPs with applied uniaxial tension to the top layer.

unreasonable to expect moiré reconstruction to become absent suddenly. The interplay between the in-plane elastic energy and interlayer vdW energy is still expected to contribute to reconstruction at higher angles due to the same fundamental physics. Although large angle TBGs have been extensively studied in the recent past, the effect of reconstruction in their structures has not been thoroughly studied.[18,19] Nevertheless, its extent remains unknown due to the current limitations of experimental methods.

Recent experimental studies have demonstrated the ability to control TBGs with and without strain and characterize moiré reconstruction for smaller $\theta$ systems.[7,13,20−24] Imaging techniques such as scanning tunneling microscopy (STM) and transmission electron microscopy (TEM) become challenging when the feature size becomes comparable to its resolution. As the size of MP decreases with an increasing twist, imaging for $\theta > 2°$ systems becomes unfeasible.[10,25] Therefore, the current understanding of reconstruction through experimental visualization is limited to low angle twists and is primarily based on image analysis techniques rather than physically measurable quantities. Aside from the scanning probe techniques, atomic reconstruction in twisted 2D heterostructures has been examined through angle-dependent scanning electron microscopy (SEM),[26] dark-field transmission electron microscopy (DF-TEM),[14,27,28] electron diffraction,[29] Bragg interferometry based on four-dimensional scanning transmission electron microscopy (4D-STEM),[20] and annular dark-field (ADF) STEM.[30,31] Such techniques have examined twisted bilayers without strain that range from close to 0° to 4° twist angles between the 2D layers. Moreover, some of these works have also examined twisted bilayers with external heterostrain, where substantial atomic reconstruction has been observed regardless of their small moiré periodicities. Such state-of-the-art electron microscopy has uncovered a tremendous amount of information about both reconstruction and how heterostrain modifies the domains in these twisted 2D heterostructures.

Optical procedures such as Raman spectroscopy offer an expedient method to characterize TBGs irrespective of their size and twist angle.[19,32−34] Still, such methods predominantly extract the collective behavior of TBGs spanning numerous MPs. Therefore, the global vibrational behavior obtained by Raman spectroscopy cannot be directly used to infer stacking and the extent of reconstruction without an interrelation of phonon behavior between local subdomains and the bulk of TBG. Atomistic analyses offer an alternative tool to study atomic arrangements locally with a fine resolution and allow for tracking of atomistic evolution with varying twist angle.[10,21,35−37] Previous works are heavily concentrated at or below the magic angle and do not explain the correlation between the local and the global behavior of TBGs. Moreover, these works have not studied the evolution of MPs with strain. As a result, there remains an outstanding question about the presence of reconstruction and its extent at higher angles, how the MP evolves with external strain, and how local and global vibrational properties are correlated.

In this work, we utilized a combination of first-principles and molecular statics atomistic simulations to examine the local domains in TBGs and how global vibrational behavior is tied to changes in local atomic registries. Based on physical parameters that include interlayer spacing and interlayer energy, our method associates each atom with known stacking types of the constituent bilayer graphene, calculates their resultant area fraction, and traces the evolution of local subdomains to demonstrate evidence of moiré reconstruction for larger $\theta$ TBG systems. This paper presents a set of criteria for identifying local stacking and reconstruction phenomena in TBGs that are valid with or without the application of strain. Additionally, we demonstrate the correlation between local and global vibrational characteristics of TBGs and discuss how it validates our results on reconstructed structures, especially at higher angles. The methods presented in this paper are devised for graphene, but further adaptations are possible for other 2D materials.

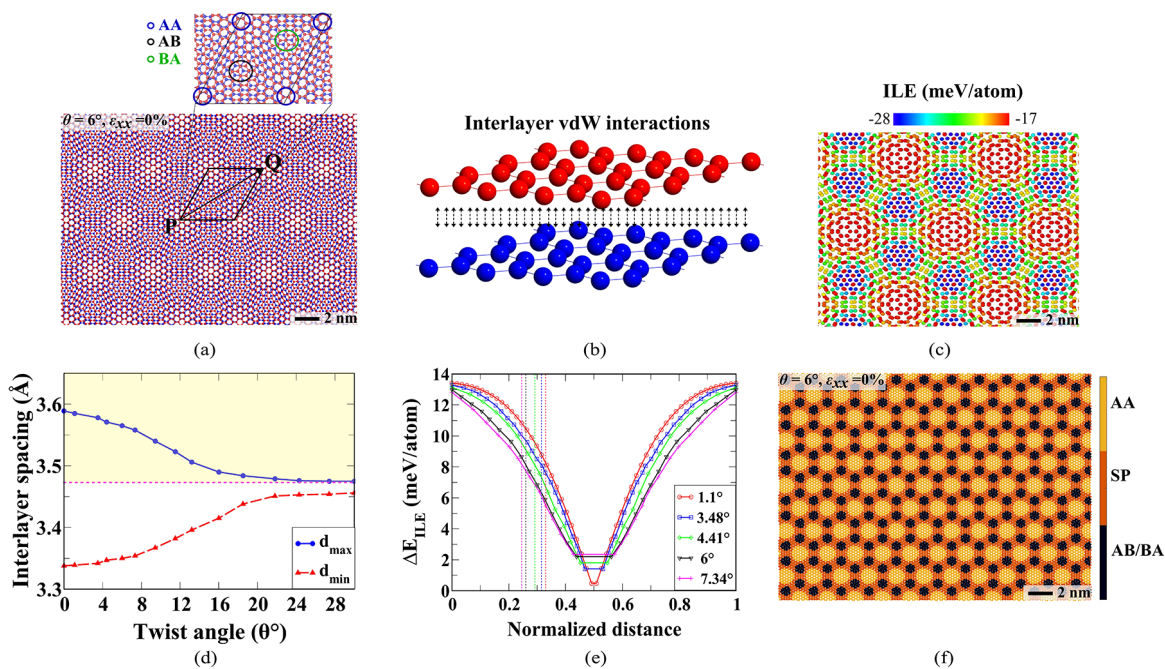

**Figure 2.** Local stacking identification method. (a) Path PQ along the center of one moiré pattern to the other ($\theta = 6°$). An inset of the real space moiré lattice is shown to illustrate the AA, AB, and BA domain centers. (b) A schematic to demonstrate the interlayer energy (ILE), which is the energy contribution of vdW interactions. (c) ILE contour plot for unstrained $\theta = 6°$ system. (d) Variation of interlayer spacing (ILS) with respect to moiré twist angles; horizontal dotted line (magenta) shows the minima of maximum ILS ($d_{max}$) obtained throughout a span of low and high angle TBGs. (e) Variation of ILE difference ($\Delta E_{ILE}$) for five representative $\theta$ values (the dotted line shows the energy difference at the soliton width boundary) (f) Contour plot demonstrating individual stacking type locally, obtained after implementing the classification method. Scale bars for the real space lattice and contour plots are shown with thick black lines.

## II. METHODS

### II.A. Atomistic Modeling

All the TBG structures are constructed by rotating the top layer of Bernal stacked bilayer graphene with respect to the bottom layer. The moiré lattice is created by identifying a common periodic lattice for the two layers. Using the TBG commensurability conditions, we have modeled their real and reciprocal space lattice parameters.[38,39] The $\vec{q}$ vector or reciprocal lattice parameter of the TBG moiré lattice is given as $\vec{q} = \vec{b'} - \vec{b}$, where $\vec{b}$ and $\vec{b'}$ denote the reciprocal lattice vectors of the bottom layer and rotated top layer, respectively. When heterostrain is applied, the strained $\vec{q}$ vector is expressed as $\vec{q_i^\varepsilon} = \vec{b^\varepsilon} - \vec{b}$, where $\vec{b^\varepsilon}$ denotes the strained top layer. The mathematical expressions of $\vec{b^\varepsilon}$ are presented in Supporting Information (SI) Section II. All the atomistic models are relaxed using density functional theory (DFT) simulations, except for the $\theta_m = 1.08°$ system. Because of a large moiré lattice for this structure (11164 atoms), DFT becomes prohibitively inefficient, so we use force-field potentials to relax this structure.

### II.B. DFT Calculations

The real space lattices of TBG systems were constructed using the ATOMISTIX TOOLKIT (QuantumATK) package.[40] All the first-principles simulations were conducted with generalized gradient approximation (GGA) assimilated in the Quantum Espresso open source package.[41,42] The Perdew−Burke−Ernzerhof (PBE) form and GGA have been used as the exchange-correlation functional.[43] Ion-electron interactions for carbon atoms in TBGs have been described by ultrasoft pseudopotentials.[44] All technical details about DFT parameters are given in SI section I.

### II.C. MS Simulations

Molecular statics (MS) simulations were done using LAMMPS open source software.[45,46] The unstrained, DFT relaxed TBG moiré lattice was transformed into an orthogonal cell for performing MS simulations. The simulation box is considered with free surface boundary conditions in the direction of uniaxial strain, allowing us to account for the aperiodic crystal geometry (or moiré lattice mismatch) due to strain applied to one of the TBG layers. The uniaxial strain was incremented by ±0.1% up to the final magnitude of ±1%. The snapshots of the structure at different strain magnitudes were taken in the Ovito open visualization tool. Further computational details are mentioned in SI section I.

## III. RESULTS AND DISCUSSIONS

### III.A. Global Structural Analysis of Pristine and Strained TBGs

We have studied a number of TBG systems between $\theta = 1.08°$ and $13.2°$ to perform our analysis on MPs close to $\theta_m$ as well as at much larger angles. For simplicity, most of the presented data analysis includes three representative TBG systems at $\theta = 1.08°$, $6°$, and $13.2°$. Figure 1a shows the MP geometries modeled using the well-defined commensurability conditions of TBG systems and relaxed using first-principles or force field-based relaxation techniques (see Methods). It must be noted that the studied TBGs systems are commensurate bilayer models with a specific rotation angle, as derived from mathematical expressions of their $\vec{q}$ vectors (SI Section II). Although TBG systems can be experimentally fabricated with an arbitrary twist angle, both commensurate or incommensurate, the synthesized crystal relaxes to a lattice approximately resembling a commensurate twist angle structure.[14,47] Geometrically, a moiré lattice with an incommensurate rotation angle results in a structure with infinite periodicity. For atomistic simulation, studying such large structures becomes

C

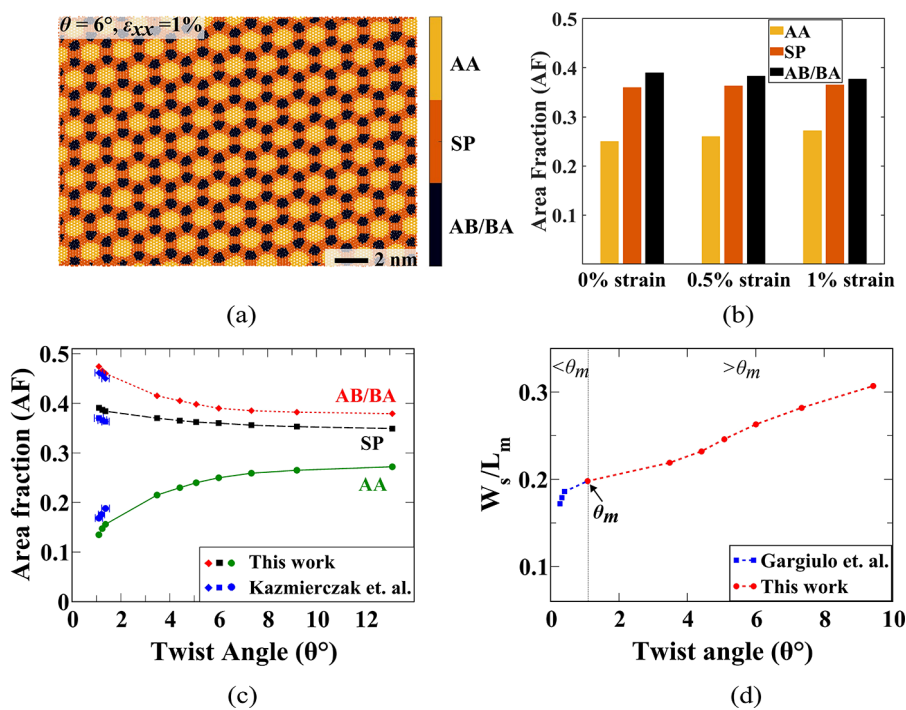

Figure 3. Evolution of local regions with twist angle and strain. (a) Contour plot demonstrating local stacking type for the heterostrained $\theta = 6°$ system (1% tension). The scale bar for the contour plot is shown with a thick black line. Area fractions of individual stacking domain with respect to (b) strain (tension) and (c) twist angle. The blue markings in (c) are extracted from reported work by Kazmierczak et al.[20] to compare our results with data obtained by analyzing experimental measurements. The error bars shown are directly inserted from the cited article. (d) Width of SP regions or soliton width ($W_s$) normalized with the length of the moiré lattice ($L_m$) as a function of the twist angle. A dotted line is drawn at the magic angle ($\theta_m = 1.1°$) to distinguish the regions below and above $\theta_m$.

computationally intractable; therefore, we have restricted our analyses to TBGs with commensurate twist angles only.

Since the local domains in TBG evolve through high symmetry BLG stacking, we can observe topographical variation in the structure[48,49] represented by the interlayer spacing (ILS) contour plot (Figure 1c). The centers of hexagonal MPs have regions of atoms where AA stacking exists.[12,50] These central regions are surrounded by two domains, AB and BA stacking, which are energetically degenerate but topologically inequivalent. Since both types of stacking represent the Bernal graphene, they can be considered to be the same category.[51,52] The boundaries of these AB/BA regions are separated by segments referred to as strain solitons caused by the shear strain due to two inequivalent stacking domains facing each other. Strain solitons have a characteristic width referred to as the soliton width.[50] The atomic structure in the soliton regions corresponds to SP stacking, an intermediate configuration between AB (or BA) and AA. A TBG system displays an out-of-plane corrugation in its structure caused by local ILS variation, with AA regions having the highest spacing followed by SP and AB regions.[10,11,36]

By applying heterostrain, we observed a similar topographical feature with distorted MPs due to the inequivalence of strain in each layer that resulted in an oblique moiré arrangement[7] (Figure 1b,d for tension and Figure S1 for compression). Geometric analysis is conducted to analyze the angular change due to distortion and rigid rotation (Figure 1e) by deducing the expressions of their reciprocal lattice ($\vec{q}$) vectors (see SI Section II). The change in the $\vec{q}$ vector with uniaxial strain triggers the distortion in MPs.[33,53] As shown in Figure 1e, the boundaries of MPs resemble a hexagon. On connecting the centers of adjacent MPs, we can draw a triangle ($\Delta ABC$) with $\overrightarrow{AB}$ and $\overrightarrow{BC}$ as the moiré lattice vectors and $\alpha$ as the angle between them. In the unstrained condition, the magnitude of vectors $|\overrightarrow{AB}| = |\overrightarrow{BC}| = L_m$ ($L_m$ = Length of MP), and the angles are $\alpha = 60°$ and $\phi = 120°$. As the $\vec{q}$ vector changes with uniaxial heterostrain, $\Delta ABC$ transforms to $\Delta A'B'C'$ such that $|\overrightarrow{A'B'}| \neq |\overrightarrow{B'C'}|$. A change in $\alpha$ can quantify the deformed moiré lattice due to the applied strain (Figure S2). The expressions of moiré reciprocal lattice vectors show the geometrical changes caused by heterostraining these systems (SI Section II).

### III.B. Classification Method to Identify Local Domains

The deformation of MP with strain gives rise to changes in their local subdomains, and it is crucial to examine them to quantify their contribution to global physical behavior. Traversing along the diagonal of MP (path PQ in Figure 2a), i.e., from the center of one moiré pattern to the center of its second nearest neighbor, we expect to cross all of the locally stacked regions: AA, AB, SP, BA, and AA.[10,50,52] Since we aim to develop criteria to classify each atom into one of these stackings, we first examined the atoms along the path PQ. To perform the stacking identification, we initially used the ILS parameter $d$ because the local domains in TBGs have interlayer distancing variations. Since pristine BLG stacking follows an increasing ILS trend from AB to SP and finally the AA region, $d_{max}$ (maximum ILS) and $d_{min}$ (minimum ILS) in TBGs can be respectively understood as the ILS of the AA and AB regions. By examining the range of ILS ($d_{max}$ and $d_{min}$) over different possible twist angles (Figure 2d), we identify the minimum value of $d_{max}$ (3.475 Å) and classify atoms above this ILS threshold as AA. It should be noted that this method does not misclassify AB and SP because this threshold is well above the ILS of pristine AB (3.33 Å) and SP (3.38 Å). Due to the small ILS difference between AB and SP, the same ILS parameter cannot be used to identify the rest of the stackings.

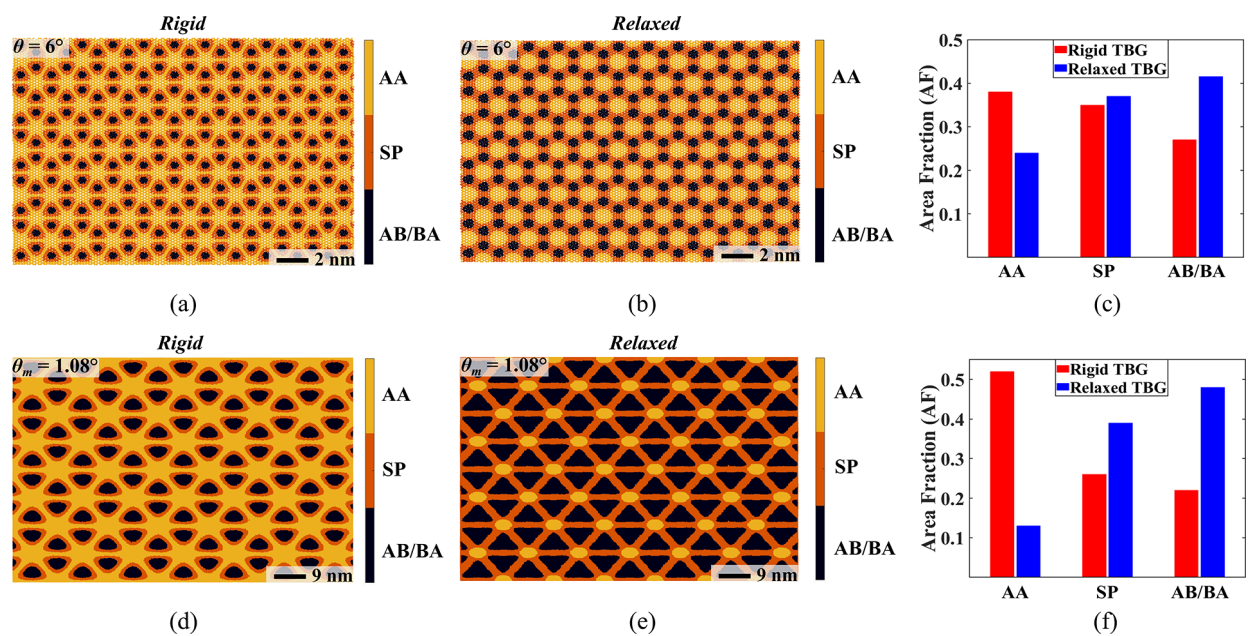

**Figure 4.** Demonstration of moiré reconstruction. Stacking contour plot for rigid, (a) $\theta = 6°$ and (d) $\theta_m = 1.08°$, and relaxed, (b) $\theta = 6°$ and (e) $\theta_m = 1.08°$, TBG systems. Scale bars for the respective contour plots are shown with a thick black line. (c, f) Comparison of area fractions for each stacking, showing the change in local atomic registries before and after relaxation that signifies the extent of reconstruction.

We introduced another parameter called "interlayer energy" (ILE) to distinguish between AB and SP according to their energy rather than ILS. The ILE is a physical measurement of vdW interaction between atoms in two different layers, as illustrated by the schematic in Figure 2b. It is obtained by computing the vdW part of the total potential energy between C atoms in different layers Figure 2c. Since these local domains have indistinguishable and strong in-plane covalent bonds, their total energy is predominantly sourced from in-plane interactions, meaning little contribution is from the interlayer interaction of atoms. Moreover, with applied strain, the changes in total potential energy due to stretching and compressing the in-plane bonds are orders of magnitude higher than the interlayer vdw counterparts. This motivates the use of vdW interaction energy and its variation for identification purposes rather than the total energy. However, being a per-atom quantity, there are fluctuations in ILE magnitudes, most prominently observed in AB regions (Figure 2c). If the average ILE magnitude is used with respect to their bonded neighbors, it will result in an insignificant difference between AB and SP subdomains. To account for this, we calculated the average ILE difference ($\Delta E_{ILE}$) of each atom with its bonded neighbors. Although separating AA and SP regions can be challenging since they have minimal fluctuations in ILE, this parameter easily classifies AB stacked atoms as they have the highest variations in energy with their neighbors. Based on the $\Delta E_{ILE}$ analysis for five representative TBGs (Figure 2e), we have identified the $\Delta E_{ILE}$ threshold at the soliton boundary (SP width) and classified atoms above that threshold (8.24 meV/atom) as AB. The infinitesimal difference in these thresholds allowed us to define a $\theta$-independent $\Delta E_{ILE}$ value for identifying the two stackings (see SI Section III for details). It is important to note that the same approach can be used for classification in the presence of strain because the physical parameters used herein do not depend on strain. Although the magnitude of interlayer energy can be expected to vary, we observed a negligible change in the $\Delta E_{ILE}$ threshold with strain (see SI Section III). Therefore, using these criteria based on ILS and ILE, we classify TBG atoms into their local stacking as shown in Figure 2f, which applies to TBGs with any twist angle and the presence of strain (Figure 3a).

Implementing the classification method, we obtained area fraction (AFs) of each subdomain present in a TBG structure. Using this measurement to monitor the evolution of local domains in the presence of strain, we observed that the subdomain AFs remain almost unchanged (Figure 3b, Figure S4 for tension and Figure S5 for compression). This demonstrates a characteristic tendency of local regions in TBG to retain their registry with an external strain applied globally. The variation of AFs as a function of the twist angle (Figure 3c) shows that area fractions of AB ($AF_{AB}$) and SP ($AF_{SP}$) increase whereas that of AA ($AF_{AA}$) decreases with decreasing $\theta$. This can be attributed to the potential energy of the soliton (SP) regions contributing to in-plane forces that displace atoms to maximize the area of AB/BA (the most stable BLG-stacking) local domains.[36] Such observations are well-studied in experiments, particularly for systems close to $\theta_m$ (1.08°). Hence we compared our theoretically estimated AFs for $\theta_m = 1.08°$ (and additional $\theta = 1.21°, 1.37°$) systems with experimentally interpreted area fractions from graphical analysis of STM images,[20] as marked in Figure 3c. The close similitude between these sets of area fraction values provides a validation of our stacking classification method. We believe our approach interprets the physical behavior of subdomains at the atomic level and with high accuracy. Furthermore, since our method is based on physical parameters such as energy, it directly encapsulates the underlying physics. In contrast, previously reported data rely on a graphical interpretation of the gradient in image intensity and contrast from experiments. Hence, our methodology is more accurate and can provide atomistic insights even at a higher twist angle where the moiré cell size shrinks drastically.

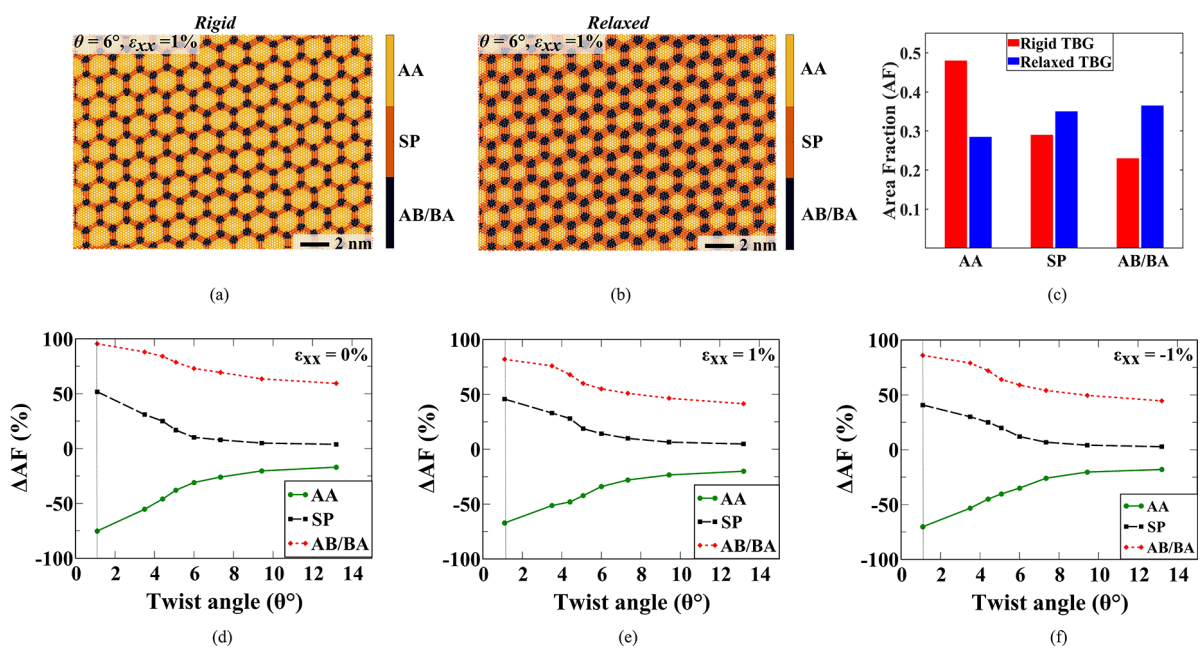

**Figure 5.** Moiré reconstruction in heterostrained TBGs. Stacking contour plots of (a) rigid and (b) relaxed $\theta = 6°$ structure in the presence of 1% uniaxial tension. Scale bars for the contour plots are shown with thick black lines. (c) Change in local stacking area fractions before and after relaxation for the strained structure. Percentage change in local $AF$s of rigid and relaxed $\theta = 6°$ structures ($\Delta AF$) with respect to twist angle for (d) pristine (unstrained), (e) 1% strained (uniaxial tension), and (f) −1% strained (uniaxial compression) TBGs. Positive and negative values of $\Delta AF$ (%) respectively indicate an increase and decrease in respective local AF. A dotted line is drawn at the magic angle ($\theta_m = 1.1°$) to distinguish the regions below and above $\theta_m$.

### III.C. Detecting Moiré Reconstruction in High Twist Angle TBGs

We further utilized this method to study atomic reconstruction in TBG systems. Moiré reconstruction can be studied by examining local regions in rigidly twisted (R-TBG) structures and comparing with their relaxed geometry.[12−15] The rigidly twisted TBG refers to its unrelaxed geometry, considered a conceptual, intermediate configuration in which the layers of BLG are twisted by a certain angle but the atoms are not allowed to reconfigure to form their true equilibrium structure. During reconstruction, local sites in the structure prefer to diverge from energetically unfavorable AA stacking by atomic displacements. This is achieved by rearrangement of the atoms to minimize vdW energy and obtain the nearly commensurate Bernal-stacked (AB/BA) BLG structure partitioned by the SP segments after reconstruction. The emergence of soliton (SP) domains is one of the predominant features of reconstruction phenomena in 2D materials. Previous studies have attributed the minor atomic displacements of relaxed large $\theta$ TBGs to an insignificant change in the atomic registry of local domains, indicating the absence of reconstruction.[14−16,54] However, examining TBG systems from an atomistic perspective and employing our subdomain identification method, we show considerable changes in the local registries for large $\theta$ TBGs. We used the area fraction measurement to capture the structural changes in local domains of relaxed and unrelaxed geometries. The stacking identification assessment of R-TBG is conducted similarly to the relaxed TBG (see SI Section IV). For the $\theta = 6°$ structure (Figure 4a−c), the AA regions shrink upon relaxation, and conversely, the AB/BA regions expand to approximate triangular domains. Undoubtedly, this structural change was expected and prominently observed for the $\theta_m = 1.08°$ system (Figure 4d−f), but we encountered a similar observation for a large-$\theta$ structure as well. Hence, contrary to the general idea that reconstruction diminishes at higher angles, we show clear evidence demonstrating moiré reconstruction in higher $\theta$ (>2°) TBG systems. This observation indicates that, irrespective of how small the atomic displacements are, the change in $AF$s of local domains for higher $\theta$ TBGs shows pronounced variation in atomic registries upon relaxation. In this work, we want to emphasize that atomic reconstruction is realized by simulating TBG models constructed by mathematical formulations that have negligible internal strain and are defect-free. However, for synthesizing TBGs, chemical vapor deposition (CVD)[18,19] and mechanical exfoliation are the two primarily used methods. Some of the first works on imaging atomic reconstruction in TBG heterostructures were done on TBGs grown via CVD[27,28] and closely resemble the reconstruction effects examined with high-quality van der Waals heterostructure assembly.[14,20] To our knowledge, direct experimental comparisons are unavailable based on the fabrication method. However, atomic relaxation and extent of reconstruction differences could still be possible due to variations in defect density, thermodynamic processes, and the nature of process-induced strain application.

### III.C.1. Comparison of Reconstructed Structures below and beyond the Magic Angle.

As the twist angle approaches perfect commensuration (0°), atomic reconstruction effects become more prominent and complex within the moiré superlattice. Atomic reconstruction produces intralayer shear strain in the system, where shear-induced displacement causes the formation of soliton walls between AB/BA domains. It has been reported that sharp domain boundaries emerge locally, separating different stacking regions in TBGs below $\theta_m$.[14] These sharp soliton boundaries or domain walls are identified as SP regions in our local domain classification method and the contour plots. The soliton boundaries appear as channels of thin rectangular shape in the reconstructed crystal structure as shown in Figure 4e. For TBG systems above $\theta_m$, the induced local lattice distortions are reduced, and soliton boundaries decrease in width. However, the moiré cell size also reduces inversely with increased twist angle. If we compare the width of solitons (identified as the SP structure type) relative to the moiré cell size as shown in Figure 3d, it is clear that the soliton boundary formation in the reconstructed TBGs beyond $\theta_m$ is still significant. Though above $\theta_m$, the shape of the soliton walls is geometrically less regular in comparison as shown in Figure 4b due to softening of the domain boundaries.[29]

For small twist angle TBGs at or below $\theta_m$, the energetically unfavorable AA domains shrink drastically upon reconstruction, which translates to a large change in their area fraction. For example, the area fraction of AA regions in the $\theta_m = 1.08°$ structure shows a dramatic reduction of 76% upon relaxation (Figure 4f). As a result, there is more space for AB/BA regions to form sharp triangular boundaries within the moiré cell. Whereas for large angle TBGs above $\theta_m$, as the moiré cell size becomes comparable with the AA and SP domain size, there is less shrinkage of AA regions (35% for $\theta = 6°$) in the reconstructed lattice. As a result, the domain changes are more gradual, and their boundaries appear less sharp. In an experimental setup, it becomes extremely difficult to resolve these different domain walls for TBGs beyond $\theta_m$ due to limitations in experimental resolution. Though atomic reconstruction has been mostly believed to become negligible in larger twists, our work indicates that the structural effect of reconstruction still exists and that there is evidence for soliton formation. Moreover, recent experimental measurements using hyperspectral imaging of exciton confinement within a moiré unit cell with a subnanometer electron probe hints toward our findings of soliton formation beyond $\theta_m$.[31]

### III.D. Analyzing Extent of Reconstruction in Strained and Unstrained TBGs

Using this approach, we have also studied moiré reconstruction in high angle TBGs in the presence of heterostrain. Lattice deformation due to heterostrain induces distortion in MPs, which is minimized by sustaining the formed domain-wall-like boundary lines (SP regions) due to superlattice reconstruction.[14,23,24] Similar to the unstrained case, we have compared the local $AF$s of rigid and relaxed systems under heterostrain (Figure 5). The rigid system for strained TBGs refers to its unrelaxed structure obtained after employing strain to the relaxed geometry of the pristine TBG structure (see SI Section IV). We observed that our assessment could capture the variations in the local atomic registry of strained TBGs (Figure 5a–c). The substantial change in $AF$s of AA and AB regions and perpetuity of SP domains signify the tendency of preserving the SP boundaries with change in the local atomic registry of AA and AB domains, thus indicating the presence of atomic reconstruction in large $\theta$ strained TBG systems. To assess the extent of change in local registries, we have calculated the percentage change in local $AF$s upon relaxing the structures, i.e., $\Delta AF(\%) = \left(\frac{AF_{relaxed} - AF_{rigid}}{AF_{rigid}}\right) \times 100$. On examining the variation of $\Delta AF$ over unstrained (Figure 5c) and strained (tensile Figure 5e and compressive Figure 5f) TBGs spanning a wide range of twist angles, it is observed that $\Delta AF$ for all local stackings monotonically decreases with increasing $\theta$. Although this implies that, as expected, the effect of reconstruction reduces with increasing twist angle, $AF$ data shows that it cannot be disregarded. For both pristine and strained cases, the AB stacked domains show ample variation in rigid and relaxed configurations, even for higher angles. This variation rapidly decreases for AA and SP regions, especially at very high twist angles. Nonetheless, this analysis reveals the existence of local atomic reconstruction for both unstrained and strained large $\theta$ TBG systems.

It has been previously argued that, for a large twist angle, the gaining vdW energy cannot compensate for the decreasing intralayer elastic energy.[14,16,24] This results in no change of vdW stacking energy between rigid and relaxed structures, ultimately indicating the absence of reconstruction. However, our analysis of ILE over different $\theta$ values (Figure S6) clearly shows a small but relatively significant difference between the rigid and relaxed structures of higher $\theta$ TBGs. Although we observed a quick increase and gradual decrease in energies of relaxed and R-TBG respectively, with increasing $\theta$, the relaxed (or reconstructed) system has the lower energy throughout. Thus, even for large twist angles, the reconstructed structure formed as a consequence of local atomic changes in their energetically favorable configuration, which directly establishes the presence of reconstruction. It is not surprising that such minor changes in atomic registries for large twist angles are challenging to capture with experiments given length scale limitations. But based on our results, structural reconstruction should not be neglected for higher angles and motivates the study of the implications of reconstruction for large-$\theta$ TBGs. Theoretical works previously suggested that atomic reconstruction effects be negligible on the electronic band structure in TBG beyond 3°,[10,11] and they subsequently have not accounted for this effect when calculating the properties of heterostrained high-angle TBGs.[7] However, when combined with strain, reconstruction in TBG structures has been experimentally investigated[55,56] and theoretically predicted to produce strongly correlated and topological states similar to that of its magic-angle counterpart.[9,23] At the magic angle, small amounts of uniaxial heterostrain are anticipated to induce quantum phase transitions between Kramer's intervalley-coherent insulator and a nematic topological semimetal.[55] Away from the magic angle, topologically nontrivial flat bands can be readily modulated with heterostrain parameters (e.g., magnitude, direction) and valley polarization states.[23] Thus, correlated electronic phases and topological states may be equally accessible in higher-angle TBG structures. We believe first-principle based calculations could help predict such heterostrain-induced phenomena upon properly accounting for atomic relaxation. Though strongly correlated and topological states are not predicted to be in TBG beyond the magic angle (without heterostrain), our work calls for a thorough investigation of all twisted 2D heterostructures with respect to both atomic relaxation and heterostrain application.

### III.E. Mapping Local and Global Physical Property (Phonon Behavior) to Changes in the Local Atomic Registry

Further validation on the presence of reconstruction at high angles lies within the interrelation of local stacking domains and global vibrational properties. To accomplish this, we have studied phonon behavior which can be directly translated to

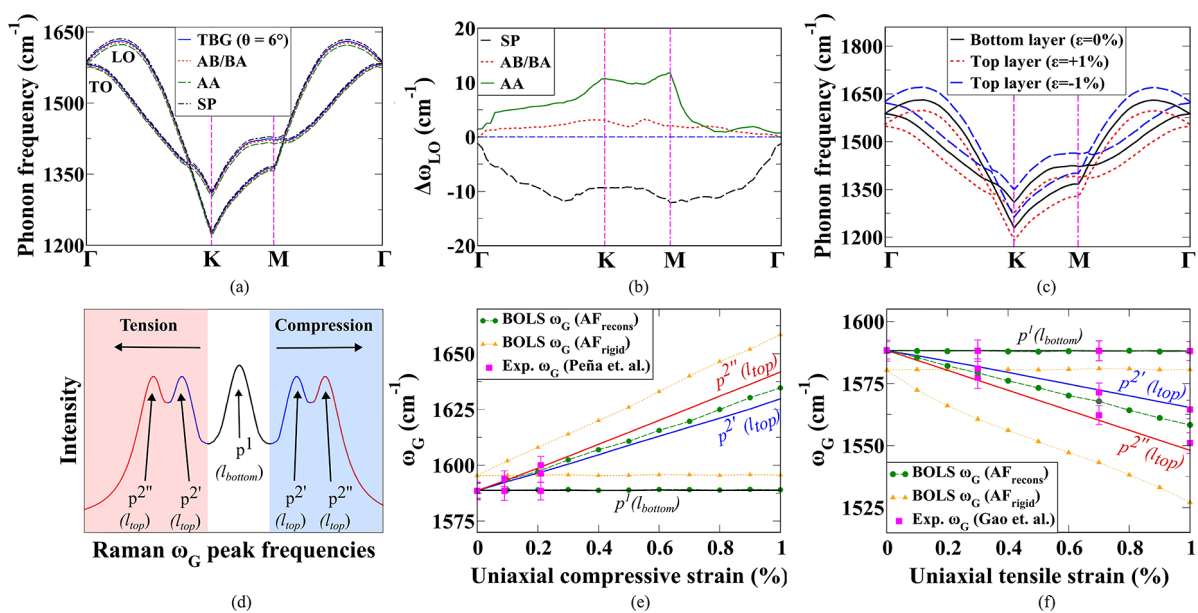

**Figure 6.** Phonon behavior of TBGs with respect to its local domains. (a) Optical phonon modes of TBLG ($\theta = 6°$) and its counterparts. (b) Longitudinal optical (LO) phonon frequency difference for each subdomain in the $\theta = 6°$ TBG system. (c) Phonon band splitting with heterostrain (tension and compression). (d) Schematic of a typical Raman G-peak splitting with inequivalent strain employed in a bilayer system. Comparison of G-band frequencies for (e) $\theta = 6°$ with uniaxial compression and (f) $\theta = 13.2°$ with uniaxial tension. Solid lines in (e) and (f) denote the Raman G-peak data obtained from DFT-based phonon calculations. Heterostrain-assisted peak splitting of the top and bottom layers (as shown in the schematic) is also denoted. Parts (e) and (f) also show the close alignment of bond order length strength (BOLS)-estimated data using reconstructed $AF$s ($AF_{recons}$) with DFT-calculated and experimental data (reported by Peña et al.[73] and Gao et al.[7]) as compared to that using rigid TBG $AF$s ($AF_{rigid}$). The error bars shown for experimental data are inserted directly from the cited articles.

Raman scattering frequencies, which is an efficient experimental technique for examining these systems, especially under strain.[57−60] We have examined the phonon dispersion spectra of TBGs and their local domains with ab initio simulations. As explained by Cocemasov et al. and Wang et al., TBG structure contains hybrid folded phonon branches resulting from different BZ directions of unrotated bilayer graphene.[19,61] The folded phonon frequencies can be obtained by zone folding of the initial phonon dispersion curve into the reduced BZ of the moiré superlattices. Using DFT, we initially calculated the phonon spectra of unstrained TBG systems and then obtained their unfolded phonon spectra (see Methods and SI Section V). Compared to the phonon spectrum of BLG, the difference in phonon modes for TBG is relatively small due to weaker interlayer interaction (Figure S7). Although we noticed some differences in low-frequency acoustic phonons, the effect is substantially feeble for optical modes that correspond to the experimentally observed Raman peaks.[61,62] Pertaining to our goal of probing Raman spectra of TBGs, we analyzed the high frequency optical (longitudinal (LO) and transverse (TO)) branches of its phonon spectra.[63] We independently computed the phonon behavior of each subdomain and compared them to the global optical vibrational behavior (see SI Section V) as shown in Figure 6a. To analyze the minute difference between phonon frequencies of all the structures, we have plotted the optical phonon frequency difference ($\Delta\omega$) of each stacking with respect to the whole TBG structure, $\Delta\omega = \omega_{TBG} - \omega_{stacking}$ (Figure 6b shows $\Delta\omega$ for LO). We observed that the AA and AB regions' phonon frequency magnitude is more petite than overall TBG, whereas it is more significant for the SP region. A similar trend is observed while comparing the TO phonon modes (Figure S8). The optical phonon behavior of AB stacking is the closest to that of TBG, indicating that AB-stacked domains predominantly control the overall phonon behavior in TBGs. This is because unfolded phonon branches of TBG exhibit an infinitesimal difference compared to that of Bernal stacked BLG.[57,62] The correlation of the $AF$ measurements with local and global phonon behavior is discussed in the following subsections.

### III.E.1. Correlating Local Area Fraction Measurement and Phonon Behavior Using Bond Order Length Strength Theory.
To further establish a connection between the optical phonons modes of TBG and phonon frequencies of its subdomains with individual stacking $AF$s, we utilized the bond order length strength (BOLS) theory.[64] BOLS can correlate Raman peaks and their shifts in constitutive structural parameters such as bond length and energy.[64−66] It explains that the intrinsic association of bonds with their physical properties can describe the extrinsic process of optical electron scattering captured by their phonon spectra. This theory provides an independent method of calculating phonon frequencies of TBG based on the $AF$s of each subdomain. Therefore, comparing the results from BOLS theory and ab initio phonon frequencies of TBG can further validate the accuracy of our subdomain categorizations. The details of BOLS formulation and the parameters involved are explained in SI Section I. To obtain the vibrational properties of various structures using BOLS correlation, we can deduce the phonon frequency shift based on bond length ($d_z$), bond energy ($E_z$), reduced mass ($\mu$), and atomic coordination number ($z$) using the following relations:

$$\Delta\omega \propto \frac{z}{d_z}\sqrt{\frac{E_z}{\mu}} \tag{1}$$

$$\Delta\omega = k\left(\frac{z}{d_z}\sqrt{E_z}\right) \quad (2)$$

$$\Delta\omega = \omega_{structure} - \omega_{bulk} = k(\beta) \quad (3)$$

where $k$ is the proportionality constant in eq 1 ($\mu$ is constant because we have only carbon-based systems). $\Delta\omega$ is the difference between the optical phonon frequency of a system and a reference material considered in bulk form (see SI Section I). Hence, $\Delta\omega = k\beta$, where $\beta$ is the prefactor containing the variable parameters, such that $\beta = \sqrt{E_z}(z/d_z)$. The magnitude of this prefactor directly relates to the optical phonon frequency of a structure $\omega_{structure}$ and thus can help in calculating its phonon behavior in terms of the associated physical parameters (i.e., $z$, $d_z$, and $E_z$). Hence, we have utilized this BOLS theory-based prefactor $\beta$ to study the phonon behavior of TBGs and their local domains, including their strained configurations.

The calculated $\beta$ magnitudes for the global TBG structure ($\beta_{TBG}$) and its subdomains are listed in Table 1, and values of all the parameters such as $d$, $z$, and $E$ are listed in a table in the Supporting Information. Although the $\beta$ magnitudes are numerically close, they follow a trend as $\beta_{SP} > \beta_{TBG} > \beta_{AB} > \beta_{AA}$, on careful inspection. This trend also aligns with the observation made while comparing the optical frequencies of these structures (Figure 6b). Interestingly, this shows how effectively the BOLS theory could endorse the characteristic trend in their phonon behavior. Furthermore, we employed the local stacking $AF$ values of reconstructed structures in the BOLS expression to instill an alternate estimation of phonon frequencies to authenticate our classification method, as will be explained. We analyzed the phonon behavior of the global TBG structure based on two approaches. The first uses $\beta_{TBG}$ calculated directly from the BOLS expression. For the second approach, we use a weighted average of $\beta$ values of individual stacking with their reconstructed $AF$s as the weights, i.e., $\beta_{TBG} = AF_{AA}\beta_{AA} + AF_{AB}\beta_{AB} + AF_{SP}\beta_{SP}$. On comparing the actual and weighted $\beta_{TBG}$, i.e., $e_{actual} = (\beta_{TBG(weighted)} - \beta_{TBG(actual)})/\beta_{TBG(actual)}$, we observed that they align very well with a small error percentage, including for strained systems (Table 2).

Table 1. List of $\beta$ ($eV^{1/2}$ $Å^{-1}$) Prefactor Values

| Stacking | $\theta = 1.08°$ | $\theta = 6°$ | $\theta = 13.2°$ |
|---|---|---|---|
| AA | 3.084 | 3.198 | 3.418 |
| AB | 3.126 | 3.294 | 3.450 |
| SP | 3.180 | 3.376 | 3.491 |
| TBG ($\beta_{BOLS}$) | 3.135 | 3.306 | 3.474 |
| TBG ($\beta_{weighted}$) | 3.141 | 3.292 | 3.466 |

Table 2. Error Table for BOLS-Estimated $\beta$ Prefactors Based on Actual and Weighted $\beta_{TBG}$, for Systems with and without Strain

| | $\theta = 1.08°$ | | $\theta = 6°$ | | $\theta = 13.2°$ | |
|---|---|---|---|---|---|---|
| Strain (%) | $e_{actual}$ | $m_i$ | $e_{actual}$ | $m_i$ | $e_{actual}$ | $m_i$ |
| 0 | 0.38 | 6 | 0.27 | 5 | 0.22 | 5 |
| 0.2 | - | - | 0.42 | 5 | 0.29 | 4 |
| 0.5 | - | - | 0.60 | 5 | 0.35 | 4 |
| 0.7 | - | - | 0.51 | 5 | 0.49 | 4 |
| 1 | - | - | 0.69 | 5 | 0.45 | 3 |

However, given the seemingly small difference in $\beta$ values of the structures, it may be argued that these small errors are not intriguing. Therefore, we have analyzed both $\beta_{TBG}$ values using a times improvement basis ($m_i$). Using this, we compared the weighted $\beta_{TBG}$, first by taking our calculated local reconstructed $AF$s as the weights and second by randomly assigning equal $AF$s (33.33% weight for three regions) to each stacking. We calculated their error percentage with actual $\beta_{TBG}$ and obtained the relative error comparison or times improvement. The $m_i$ values in Table 2 show significant times improvement on considering our estimated $AF$ values of reconstructed structures. The similitude between global $\beta_{TBG}$ and weighted $\beta_{TBG}$ using local $AF$s signifies that the physical attributes of local regions in a TBG structure directly correlate with the global vibrational comportment. This analysis also reinforces that our stacking classification is an effective method to detect reconstruction in TBG structures with wide-ranging $\theta$ and strain magnitudes.

**III.E.2. Comparison of BOLS-Estimated Phonon Frequencies with Experimental Raman to Validate Subdomain Area Fraction measurement.** To validate our reconstructed $AF$ measurements with DFT-based phonon calculations and $AF$ driven BOLS theory, we first calculated the phonon spectra of strained TBGs using DFT simulations followed by calculating Raman frequencies using BOLS (see SI Section I). Figure 6c shows the optical phonon branches of TBG ($\theta = 6°$), including tensile and compressive uniaxial heterostrain. We have considered the Raman G band frequency in this study, which can be obtained at the $\Gamma$ point in the high symmetry Brillouin Zone (BZ) path.[63,67] We observed strain-induced phonon band splitting due to inequivalent strain present in both the layers[67−70] (SI Section VI). This phenomenon is observed in Raman spectroscopy as represented by the schematic of G-band Raman peaks in heterostrained TBGs (Figure 6d). Due to weak interlayer vdW interaction in TBGs, their interlayer shear strength is negligible, resulting in slippage between the layers. Hence, the bottom layer remains mostly unstrained when straining the top layer.[70,71] The Raman spectra of heterostrained TBG show significant individual peaks of the unstrained bottom layer ($p^1_{\varepsilon=0}$) and strained top layer ($p^{2\prime}$). The peak of the strained layer redshifts or blueshifts depending on the nature of strain. Also, for the case of graphene, an increase in the magnitude of strain further splits the G-band peaks corresponding to the doubly degenerate $E_{2g}^+$ and $E_{2g}^-$ phonons ($p^{2\prime\prime}$ in Figure 6d−f).[7,72]

We then used the local $AF$ values of reconstructed systems in the BOLS expression to estimate Raman G-band frequencies for comparison with experiments and establish a connection between global and local vibrational behavior. We first extracted the G-band frequency ($\omega_G$) from DFT-simulated phonon spectra for both unstrained and strained structures. Figure 6e,f respectively shows the variation of $\omega_G$ for 6° and 13.2° with strain. To demonstrate both directions of uniaxial strain, we showed the case of compression for 6° and tension for 13.2°. In both cases, we observed that $\omega_G$ at zero strain is 1588 cm$^{-1}$, which changes negligibly for the unstrained bottom layer. In Figure 6e, due to compression, we observed a blueshift in $\omega_G$ and redshift for tensile strain in Figure 6f (see SI Section VI). On comparing our results for 6° and 13.2° systems with the experimental data reported by Peña et al.[73] and Gao et al.,[7] respectively, we found good agreement between them (magenta data points in Figure 6e,f). Finally, to achieve an experimental validation of our stacking identification method as well as to highlight that the global behavior, such as Raman scattering, is tied to local structural configurations, we used our calculated AFs of reconstructed TBGs in BOLS to predict the Raman G-band frequencies of heterostrained systems (see SI Section I for details).

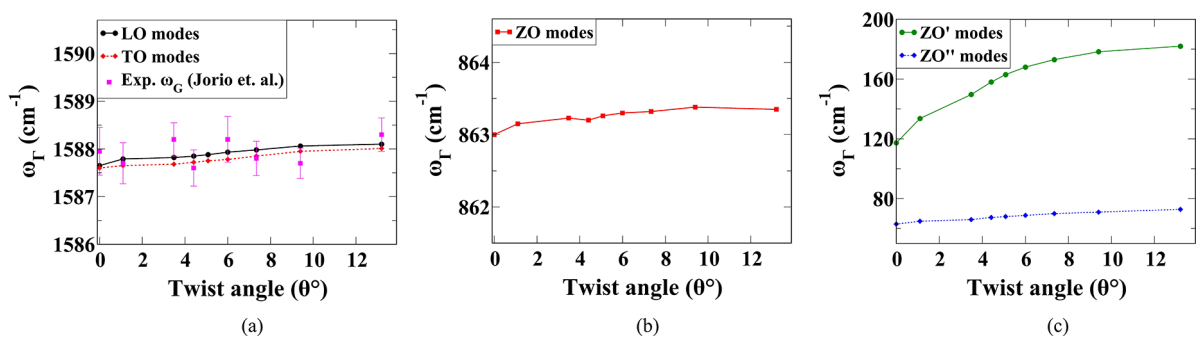

Figure 7. Comparison of optical phonon modes at the Γ point in the high symmetry Brillouin zone. (a) Longitudinal optical (LO) and transverse optical (TO) phonon frequencies ($\omega_\Gamma$) of TBGs at the Γ point as a function of the twist angle ($\theta$). The magenta points show the experimentally obtained G band peaks ($\omega_G$) pertaining to LO and TO frequencies at the Γ point, as reported by Jorio et al.[32] The error bars shown for experimental data are inserted directly from the cited article. (b) Low frequency out-of-plane phonon modes (ZO) at Γ with respect to $\theta$. (c) Low frequency layer breathing mode (ZO′) and shearing mode (ZO″) optical phonons of TBGs at the Γ point as a function of twist angle.

We found a qualitative agreement between BOLS estimated and DFT-calculated $\omega_G$ Raman peaks shown in Figure 6e,f (green dots). It must be noted that since the BOLS approach uses mathematical interpolation for projecting the phonon frequencies, it cannot resolve the further band splitting of the strained top layer. We have also used the rigid TBG AFs to check how it compares with the estimated G-band frequencies. We observed a distinct misalignment between BOLS-estimated Raman data using rigid AFs with that of reconstructed AFs and experimentally obtained data. Hence, our analysis demonstrates the difference in vibrational behavior between reconstructed and rigid structures and that the reconstructed systems align closely with the experimentally obtained measurements. This certainly implies that the physical behavior of TBGs, such as vibrational properties, is governed by the reconstructed phases even for a large-$\theta$ system, further validating the presence of moiré reconstruction in their structures. Moreover, an agreement between the AF utilized BOLS-estimated Raman data and DFT-calculated phonon shows a theoretical approach to calculate Raman frequencies at a lower computational cost. We have calculated the G-band data for the heterostrained 1.08° system using the BOLS formulation (Figure S9). Hence, using our stacking classification method and TBG Raman signature using BOLS, we demonstrated reconstruction in high twist angles and a connection between the global phonon shift in TBGs and changes in local atomic registries.

It was previously argued that interlayer coupling becomes insignificant for TBGs with large twist angles.[74] However, based on Raman spectra analysis, a couple of recent works have demonstrated that this is not the case.[19,39,61] The existence of strong folded phonons and G band resonance with an increasing twist angle hints toward effective interlayer coupling even for higher angle TBGs. This observation aligns very well with our G-band data shown as a function of the twist angle in Figure 7a. The G band peaks obtained from LO and TO phonon modes at the Γ point change very negligibly with varying twist angles. These high frequency optical phonons are not affected by the change in moiré lattice and interlayer vdW interaction in these systems. This signifies the presence of interlayer coupling in high angle TBGs that plays a major role in obtaining the in-line resonant G band frequencies.[33,75−77] This is also the case for low-frequency optical modes as shown in Figure 7b, which change insignificantly as a function of twist angle. These optical phonons represent the in-plane intralayer vibration of atoms in their lattice. In TBGs, since varying the twist angle does not significantly impact the in-plane atomic registry of each layer, we observed an intangible change in their optical frequencies. Moreover, the presence of interlayer coupling, even for large angles, results in degenerate optical phonon modes.[34,62] Although a similar behavior is noticed for the low frequency interlayer shearing modes (ZO″), the layer breathing modes (ZO′) show a dependency on the twist angle (Figure 7c). This implies that the vdW coupling plays a major role in preventing change in shearing modes for large angle TBGs. However, the incoherent out-of-phase breathing phonon modes are strongly affected by varying interlayer interaction with increasing twist angle. A similar phenomenon is observed on comparing the low frequency optical phonons of monolayer and multilayer graphene.[25,61] Although we have only discussed the phonon modes on G-band peaks in this paper, we want to emphasize the folded phonon frequencies in TBG systems that show a clear dependency on their rotation angle. This phonon peak, known as the R band, can be estimated from the reciprocal lattice vectors of the moiré lattice. It is observed to be an indispensable Raman signature used to identify the rotation angle in TBG systems.[34,35,39] We have also analyzed the LO and TO mode changes at the Γ point for each subdomain in the TBG structure. Similar to the global phonon modes and as expected, the individual local regions also show minor changes in G-band frequencies with increasing twist angle (Figure S10). Hence, our analysis signifies the importance of interlayer coupling and vdW interactions in governing the high frequency optical modes, locally and globally, for TBG systems beyond the magic angle. Moreover, the differences in phonon modes for the rigid and reconstructed structures clearly demonstrate the interplay of atomic reconstruction in large angle TBGs.

## IV. CONCLUSION

Using atomistic simulations, we studied the characteristics of locally stacked domains in TBG moiré patterns and demonstrated a comprehensive approach to study atomic reconstruction phenomena in these structures, including in the presence of heterostrain. We proposed a way to classify TBGs into their stacking types (AA, AB, and SP) and calculated area fractions of each region to track structural evolution as a function of $\theta$ and strain. Our classification scheme allowed us to show the existence of moiré reconstruction even for larger twist angle (>2°) TBG systems, which is difficult to detect experimentally. We showed how the moiré patterns of these large-angle TBGs can be distorted by applying strain. Besides, the atomic reconstruction (in terms of area fraction change of different stacking domains) can be further tuned by applying heterostrain, opening up new opportunities for large angle TBGs to be used in strain engineering applications.

We inspected TBG reconstruction over a wide range of $\theta$ and observed how it evolves in the presence of strain. To further analyze this finding and validate the area fraction ($AF$) measurement, we utilized DFT-based phonon calculations and a theoretical approach (BOLS theory) to deduce Raman frequencies and compare them with experimental data. Using BOLS theory, we discovered that global phonon behavior is directly related to the physical features of local regions. Furthermore, we realized that the Raman data using reconstructed $AF$s in BOLS aligns closely with DFT-calculated and experimental data. Comparing the Raman data with rigid AF, our results clearly differ from that of the reconstructed subdomains, implying that the latter governs the physical behavior in TBGs even for higher angles. Our study shows a self-consistent approach to characterize local regions in TBGs and utilize them to examine as well as validate moiré reconstruction phenomena based on physical measurements. The presence of reconstruction in large angle TBGs might open up an interesting avenue in the current state of twistronics research. Our methodologies can be utilized to identify stacking types and perform similar analyses in other twisted vdW systems, especially in the presence of strain.

## ■ ASSOCIATED CONTENT

### Data Availability Statement

The data supporting this study's findings are available from the corresponding author upon reasonable request.

### Supporting Information

The Supporting Information is available free of charge at https://pubs.acs.org/doi/10.1021/acsaenm.2c00259.

> Additional description of computational and theoretical methods; information on geometric analysis of TBGs, stacking identification method, and phonon spectra analysis; and additional figures for area fraction analysis of compressively strained TBGs, comparison of interlayer energy between rigid and relaxed structures, and phonon frequencies data at the Γ point for each subdomain (PDF)

## ■ AUTHOR INFORMATION

### Corresponding Author

**Aditya Dey** − *Department of Mechanical Engineering, University of Rochester, Rochester, New York 14627, United States;* orcid.org/0000-0002-5041-8662; Email: adey2@ur.rochester.edu

### Authors

**Shoieb Ahmed Chowdhury** − *Department of Mechanical Engineering, University of Rochester, Rochester, New York 14627, United States*

**Tara Peña** − *Department of Electrical and Computer Engineering, University of Rochester, Rochester, New York 14627, United States*

**Sobhit Singh** − *Department of Mechanical Engineering, University of Rochester, Rochester, New York 14627, United States*

**Stephen M. Wu** − *Department of Electrical and Computer Engineering, University of Rochester, Rochester, New York 14627, United States;* Present Address: (S.M.W.) Department of Physics and Astronomy, University of Rochester, Rochester, New York 14627, United States

**Hesam Askari** − *Department of Mechanical Engineering, University of Rochester, Rochester, New York 14627, United States;* orcid.org/0000-0001-5562-1363

Complete contact information is available at:
https://pubs.acs.org/10.1021/acsaenm.2c00259

### Author Contributions

§A.D. and S.A.C. contributed equally to this work.

### Notes

The authors declare no competing financial interest.

## ■ ACKNOWLEDGMENTS

We acknowledge the support from the National Science Foundation (OMA-1936250) and National Science Foundation Graduate Research Fellowship Program (DGE-1939268).

## ■ REFERENCES